\documentclass{ifacconf}

\usepackage{graphicx}      
\usepackage{natbib}        

\usepackage{graphicx} 
\usepackage{amsmath} 
\usepackage{amssymb}  

\usepackage{bm}
\usepackage{bbm}
\usepackage{mathtools}
\usepackage{upgreek}
\usepackage{accents}
\usepackage{enumerate}

\def\R{{\mathbb R}}
\def\N{{\mathbb N}}

\def\sign{\,\mathrm{sign}}

\def\S{\mathcal{S}}

\def\mex{\hfill$\triangle$}

\newtheorem{exa}{Example}

\newtheorem{as}{Assumption}

\begin{document}

\begin{frontmatter}

\title{Implementing prescribed-time convergent control: sampling and robustness}

\thanks[footnoteinfo]{Corresponding author: Hernan Haimovich (e-mail: haimovich@cifasis-conicet.gov.ar). Partially supported by Agencia I+D+i grant PICT 2018-1385, Argentina, and by the Christian Doppler Research Association,
the Austrian Federal Ministry for Digital and Economic Affairs and the National
Foundation for Research, Technology and Development, Austria, and by the Consejo Nacional de Ciencia y Tecnología (CONACYT-México) grant 739841.}

%

\author[HHaddress]{Hernan~Haimovich}, 
\author[Zaragoza]{Rodrigo~Aldana-Lopez}, 
\author[Austria]{Richard~Seeber}, 
\author[Intel,TECMM]{David~Gomez-Gutierrez}

\address[HHaddress]{International French-Argentine Center for Information and Systems Science (CIFASIS),\\ CONICET-UNR, Ocampo y Esmeralda, 2000 Rosario, Argentina.}
  
\address[Zaragoza]{University of Zaragoza, Departamento de Informatica e Ingenieria de Sistemas (DIIS), Zaragoza, Spain 
}

\address[Austria]{Graz University of Technology, Institute of Automation and Control, Christian Doppler Laboratory for Model Based Control of Complex Test Bed Systems, Graz, Austria 
}
  
\address[Intel]{Intel Tecnolog\'ia de M\'exico, Intel Labs, Intelligent Systems Research Lab, Jalisco, Mexico 
}

\address[TECMM]{Tecnológico Nacional de México, Instituto Tecnológico José Mario Molina Pasquel y Henríquez, Unidad Académica Zapopan, Jalisco, Mexico.}

\begin{abstract}

    According to recent results, convergence in a prespecified or prescribed finite time can be achieved under extreme model uncertainty if control is applied continuously over time. This paper shows that this extreme amount of uncertainty cannot be tolerated under sampling, not even if sampling could become infinitely frequent as the deadline is approached, unless the sampling strategy were designed according to the growth of the control action. Robustness under model uncertainty is analyzed and the amount of uncertainty that can be tolerated under sampling is quantified in order to formulate the least restrictive prescribed-time control problem that is practically implementable. Some solutions to this problem are given for a scalar system. Moreover, either under a-priori knowledge of bounds for initial conditions, or if the strategy can be selected after the first measurement becomes available, it is shown that the real, practically achievable objectives can also be reached with linear time-invariant control and uniform sampling. These derivations serve to yield insight into the real advantages that implementation of prescribed-time controllers may have.
\end{abstract}


\end{frontmatter}

\section{Introduction}
\label{sec:intro}



Recently developed methodologies for the design of observation and control algorithms under time constraints are based on time-varying gains that tend to infinity at the terminal time. These algorithms have been referred to as prescribed-time algorithms for their ability to induce observation/control error dynamics that have a prescribed settling time for every non-zero initial condition~\citep{Holloway2019,sonwan_ijrnc18}. Since the introduction of this prescribed-time control methodology by~\cite{sonwan_auto17}, such time-varying gains have been widely exploited in the literature, developing prescribed-time algorithms that exhibit remarkable properties, see e.g. the survey by~\cite{Ye2022prescribed}. For instance, the observer proposed by~\cite{Espitia2022SensorSystems} retains the prescribed-time convergence property even in the presence of sensor delay, whereas the controllers by~\cite{sonwan_auto17,ZHOU2020108760,Hua2021adaptive} retain the prescribed convergence property despite significant uncertainty (of specific structure) in the system.

It is widely acknowledged that the presence of a time-varying gain that tends to infinity makes a prescribed-time algorithm challenging to implement, see e.g.,~\citep[Section~3.2c]{sonwan_auto17} and \cite[Section~2.A]{Holloway2019}. Hence, studying the performance of such algorithms in practical settings as well as what properties are retained when workarounds are proposed is of paramount importance. For example, \citet{shakou_auto22} studies the effect of freezing the gain a short time before the deadline, a common workaround to maintain a bounded time-varying gain~\citep{Bertino2022}, and recent studies show important performance limitations in the presence of measurement noise~\citep{Aldana2022inherent}. However, to our best knowledge, the effect of sampling in the performance of prescribed-time algorithms has not been studied. 


The implementation of any control strategy in a digital processor inevitably requires sampling. The use of sampling and hold makes the closed-loop system operate in open-loop between samples. This situation can be extremely detrimental for the implementation of a prescribed-time control strategy, insofar as not just open-loop operation between samples has to be considered but also the fact that only a finite number of control actions can be taken before the deadline is reached. A theoretical way of including sampling without losing robustness, or at least minimizing this loss, would be to increase the sampling frequency over time and allow it to become infinite as the deadline is approached, causing the sampling instants to accumulate towards the deadline. This theoretical (not practically realizable) strategy could perhaps retain some of the robustness of the original continuous-time strategy. Strategies that increase the sampling frequency (without becoming infinite) as the target approaches are known to be employed by bats once a prey is echolocated and pursuit begins \citep{griweb_ab60}.

The contribution of the current paper is threefold. First, it is shown that even if sampling were allowed to become infinitely frequent toward the deadline, the sampling strategy still has to be designed in relation to the control law to retain at least some of the original robustness. Second, the amount of uncertainty that can be tolerated under sampling is quantified and a more realistic and practically achievable prescribed-time control problem is formulated. Third, a solution to this more realistic control problem is provided for the case of a scalar system. It is also shown that if the sampling strategy can be selected based on knowledge of a bound for initial conditions, then the control problem can be solved through linear time-invariant control and uniform sampling. The given results yield insight into the real advantages that implementation of prescribed-time control strategies may have.

\textbf{Notation:} $\R$, $\R_{>0}$ and $\R_{\ge 0}$ denote the real, positive and nonnegative real numbers, respectively. If $\alpha\in\R$, then $|\alpha|$ denotes its absolute value. Given a sequence $\{t_k\}_{k=1}^N$, the increments are defined as $\Delta_k := t_{k+1} - t_k$; evaluation of a function $x(\cdot)$ at $t_k$ is denoted as $x_k:= x(t_k)$.

\section{Robustness under Sampling}

Consider the scalar system
\begin{align}
  \label{eq:sc-pert}
  \dot x = f(x,t) + b(x,t) u,
\end{align}
where $f$ and $b$ are uncertain and satisfy Assumptions~1 and~2 of \cite{sonwan_auto17}, which we state here as Assumption~\ref{as:cont-time}.
\begin{as}
    \label{as:cont-time}
    There exist $\underline{b} > 0$, a known continuous function $\psi : \R \to \R_{\ge 0}$ and a locally integrable and bounded function $d : \R_{\ge 0} \to \R_{> 0}$ with unknown bound, such that
    \begin{align}
        \label{eq:b-cont}
        0< \underline{b} \le b(x,t) < \infty\\
        \label{eq:f-cont}
        |f(x,t)| \le d(t)\psi(x),&\qquad \forall x\in \R, t\in\R_{\ge 0}.
    \end{align}
\end{as}
In~\cite{sonwan_auto17}, it is shown that a time-varying state feedback $u(t)=\alpha(x(t),t)$, defined from initial time $t_0$ up to $t_0 + T$, i.e. $\alpha(x,\cdot) : [t_0,t_0+T) \to \R$, is able to drive any initial state to 0 exactly at $t_0 + T$ while tolerating the amount of uncertainty indicated by Assumption~\ref{as:cont-time}.

\subsection{The need for an upper bound on $b(x,t)$}
  Next, it will be shown that under Assumption~\ref{as:cont-time}, not only convergence but also reducing the norm of the state is impossible under sampling, even if sampling were allowed to become infinitely frequent towards $t_0 + T$. Moreover, depending on the growth of the control action towards the deadline, the state may even diverge with $b(x,t)$ bounded also from above. 
\begin{lem}
  \label{lem:inst-smpl-1}
  Consider the scalar system~(\ref{eq:sc-pert}), an initial time $t_0\ge 0$, an increasing sequence of sampling instants $\{t_k\}_{k=1}^N \subset (t_0,\infty)$, with either
  \begin{itemize}
      \item $N\in\mathbb{N}$ and $t_N = t_0 + T$, or
      \item $N=\infty$ and $\lim_{k\to\infty} t_k = t_0 + T$.
  \end{itemize}  
  Suppose that the true input remains at a constant value between sampling instants (zero-order hold), i.e.
  \begin{align}
    \label{eq:zoh}
    u(t) = \bar u_k,\quad \forall t\in[t_k,t_{k+1}),
  \end{align}
  where the control action $\bar u_k$ can depend on knowledge of the state, the bounds assumed, and even of the whole sampling-time sequence $\{t_k\}$.
  Then, for every initial condition $x(t_0) \in \mathbb{R}$ there exist functions $f$ and $b$ satisfying Assumption~\ref{as:cont-time} and such that 
  \begin{enumerate}[a)]
  \item $|x(t_k)| \ge |x(t_{k-1})|$ for all $k\in \{1,\ldots,N\}$;\label{item:nonconv}
  \item if $N=\infty$ and an infinite number of control actions is nonzero, then $\lim_{k\to\infty} |x(t_k)| = \infty$; \label{item:diverge}
  \item if $N=\infty$, $x(t_0) \neq 0$ and there exist $\overline{b} \ge \underline{b}$ and $c > 2$ such that such that $|\bar u_k| \ge c|x(t_k)|/(\overline{b} \Delta_k)$, then $\lim_{k\to\infty} |x(t_k)| = \infty$ holds in addition with $b$ upper bounded by $\overline{b}$.\label{item:bbound}
  \end{enumerate}
\end{lem}
\begin{pf}
For $c > 2$, define a sequence $\{b_k\}_{k=0}^N$ as
  \begin{align*}
    b_k = 
    \begin{cases}
      \max\left\{\underline{b}, \dfrac{c|x(t_k)|}{|\bar u_k| \Delta_k} \right\} &\text{if }\bar u_k \neq 0,\\
      \underline{b} &\text{otherwise.}
    \end{cases}
  \end{align*}
  Let $f\equiv 0$ and $b(x,t) := \tilde b(t) := b_k$ for $t\in [t_k,t_{k+1})$, and define $x_k := x(t_k)$. Integration of~(\ref{eq:sc-pert}) gives
  \begin{align*}
    x_{k+1} = x_k + \int_{t_k}^{t_{k+1}} b(x(s),s) u(s) ds = x_k + \Delta_k b_k \bar u_k \\
    = 
    \begin{cases}
      x_k &\text{if }\bar u_k = 0,\\
      x_k + \max\left\{\Delta_k \underline{b} |\bar u_k|, c|x_k| \right\} \sign(\bar u_k) &\text{if }\bar u_k \neq 0.
    \end{cases}
  \end{align*}
  As a consequence,
  \begin{align}
    \label{eq:xkineq}
    |x_{k+1}| \ge 
    \begin{cases}
      |x_k| &\text{if }\bar u_k = 0,\\
      (c-1)|x_k| &\text{if }\bar u_k \neq 0.
    \end{cases}
  \end{align}
  Items~\ref{item:nonconv}) and~\ref{item:diverge}) follow straightforwardly from these inequalities. For item~\ref{item:bbound}), it happens that
  \begin{align*}
    \max\left\{\underline{b}, \dfrac{c|x_k|}{|\bar u_k| \Delta_k} \right\} \le \max\left\{\underline{b}, \overline{b} \right\} = \overline{b}
  \end{align*}
  so that the sequence $\{b_k\}$ is clearly upper bounded by $\overline{b}$. If $x_0 = x(t_0) \neq 0$, then $x_k \neq 0$ and thus $\bar u_k \neq 0$ for all $k\in\N_0$ follows from the assumption on $|\bar u_k|$ and~(\ref{eq:xkineq}).\qed
\end{pf}
The main problem introduced by sampling is the fact that a zero crossing of $x$ cannot be instantaneously detected and acted upon. As a consequence, there is necessarily a delay between the instant when $x$ changes sign (crosses zero) and the instant when the control action can be updated to steer the state in the right direction. This delay is not present in the analysis of prescribed-time control of \cite{sonwan_auto17,sonwan_ijrnc18,holkrs_auto19}. 

Lemma~\ref{lem:inst-smpl-1} implies that there is a limitation to the kind of uncertainty that can be successfully tolerated under sampling. Specifically, some upper bound on $b$ should exist, or otherwise nonzero control actions could cause the state not only to not converge but also to diverge. Moreover, the time dependence of the control law, more precisely, the way in which the control action can increase in magnitude for fixed $x$, has to be related to the decrease in the sampling periods in a specific way as the deadline is approached, or otherwise divergence can occur even with bounded $b$.

\begin{exa}
  \label{exa:tel}
  Consider a sampled implementation of the following time-varying control law where $t_0 = 0$ and $T=1$ are selected for simplicity:
  \begin{align*}
    u &= -k(t)x, \ k(t) = \frac{A}{(1-t)^m},\ t\in [0,1), A>0, m\in\N \\
    t_k &= 1 - a^k, \quad \text{with }a \in (0,1),\\
    \bar u_k &:= u(t_k) = -k(t_k) x(t_k),\\
    \Delta_k &= t_{k+1}-t_k = a^{k}-a^{k+1} = a^k (1-a).
  \end{align*}
  For this implementation, it follows that
  \begin{align*}
    k(t_k)\Delta_k = \frac{a^k(1-a)A}{a^{km}},
  \end{align*}
  and hence
  \begin{align*}
    |\bar u_k| = \frac{a^{(1-m)k}(1-a)A |x(t_k)|}{\Delta_k}.
  \end{align*}
  According to Lemma~\ref{lem:inst-smpl-1}, even bounded uncertainty in $b$ with $\overline{b} > \frac{2}{(1-a)A}$ for $m = 1$ or with arbitrary bounds for $m > 1$ can make the sampled implementation diverge. \mex
\end{exa}
The problem illustrated in the previous example is due to the control action growing fast in relation to the frequency of updates given by sampling. Any control law with even faster growth will have the same problem (or even worse). A solution would be to design the sampling-time sequence in correspondence with the control law.
\begin{exa}
  \label{exa:sqrt}
  Consider again the control law of Example~\ref{exa:tel}, with $t_0=0$ and $T=1$, but updated as follows:
  \begin{align*}
    t_k &= 1- a^{(k^q)},\quad \text{with }a \in (0,1), q=\frac{1}{m^2+1},\\
    \Delta_k &= a^{( k^q)} - a^{ ((k+1)^q)}.
  \end{align*}
  In this case,
  \begin{align*}
    k(t_k) \Delta_k &= \frac{A}{a^{m (k^q)}} [a^{( k^q)} - a^{ ((k+1)^q)}]\\ 
    |\bar u_k| &= |u(t_k)| = \frac{A (a^{( k^q)} - a^{ ((k+1)^q)}) |x(t_k)|}{a^{m (k^q)} \Delta_k}.
  \end{align*}
  It can be shown that
  \begin{align*}
      \lim_{k\to\infty} \frac{a^{( k^q)} - a^{ ((k+1)^q)}}{a^{m (k^q)}} = 0
  \end{align*}
  and hence item~\ref{item:bbound}) of Lemma~\ref{lem:inst-smpl-1} cannot hold. \mex
\end{exa}
These examples indicate that even in the theoretical but not physically implementable case of having infinitely frequent sampling towards the deadline, the mere presence of sampling causes additional problems. 

%
%

\subsection{The need for a known upper bound on $f(x,t)$}

Consider again the scalar system~(\ref{eq:sc-pert}), a deadline $T > 0$ and a finite number of sampling instants, arranged in an increasing sequence $\{t_k\}_{k=1}^N$ with $t_N = t_0 + T$. According to Lemma~\ref{lem:inst-smpl-1}\ref{item:nonconv}), convergence under any form of sampling requires the function $b$ in~(\ref{eq:sc-pert}) to be upper bounded. Consider therefore the following replacement for~\eqref{eq:b-cont}:
\begin{align}
    \label{eq:b-disc}
    0 < \underline{b} \le b(x,t) \le \bar{b} < \infty.
\end{align}
With only a finite number of samples before the deadline, it is not possible to tolerate an unknown bound for the function $f$ in~\eqref{eq:sc-pert}, as shown next.

\begin{lem}
\label{lem:up-bnd-f}
    Consider the scalar system~(\ref{eq:sc-pert}) with $b$ bounded as in~\eqref{eq:b-disc}, a deadline $T > 0$ and a finite number of sampling instants, arranged in an increasing sequence $\{t_k\}_{k=1}^N$ with $t_0 < t_1$ and $t_N = t_0 + T$. Suppose that the control action remains constant between sampling instants, as per~\eqref{eq:zoh}. 
    Let $\epsilon > 0$ and suppose that $\psi : \R \to \R_{\ge 0}$ satisfies $\inf_{|s| \ge \epsilon} \psi(s)/|s| > 0$.
    Then, for every $M>0$, there exist functions $d:\R_{\ge 0} \to \R_{\ge 0}$ and $f$ satisfying Assumption~\ref{as:cont-time} such that $|x(t_0)| \ge \epsilon$ implies $|x(t_N)| \ge M$ for every $b$ satisfying \eqref{eq:b-disc}.
\end{lem}
\begin{pf}
    Define $a = \inf_{|s| \ge \epsilon} \psi(s)/|s|$, $c = \max\{1, \sqrt[N]{M/\epsilon} \}$, and select $f(x,t) = d(t) a x$ for $|x| \ge \epsilon$, $f(x,t) = 0$ for $|x| < \epsilon$.
    Let $\underline{\Delta} = \min_{k\in\{0,\ldots,N-1\}} \Delta_k$, set
    \begin{equation}
        D = \frac{1}{a \underline{\Delta}} \max\left\{ 2 (1+c) \frac{\overline b}{\underline b}, \ln(1+2(c-1)) \right\}
    \end{equation}
    and define $d(t) = d_k$ on each interval $[t_{k}, t_{k+1})$, with $d_k \in [0,D]$ being specified later.
    It will be shown that $|x_{k+1}| \ge c |x_k|$ for all $k =0, \ldots, N-1$, proving the claim due to the selection of $c$.
    For each $k$, define $\underline q_k = \underline{b} \Delta_k \bar u_k / x_k$, $\overline q_k = \overline{b} \Delta_k \bar u_k / x_k$, $q(t) = b(x(t),t) \Delta_k \bar u_k/x_k$ for $t\in[t_k,t_{k+1})$, and the convex combination $q_k(\xi) = (1-\xi) \underline q_k + \xi \overline q_k$.
    Then, for every integrable function $g : [t_k,t_{k+1}] \to \R_{\ge 0}$ and every $t \in [t_k,t_{k+1}]$ there exists a $\xi$ such that 
    \begin{equation}
    \label{eq:modified-mvt}
        \int_{t_k}^{t} q(\tau) g(\tau) d \tau = q_k(\xi) \int_{t_k}^{t} g(\tau) d\tau.
    \end{equation}
    Distinguish two cases: if $\underline q_k \le -(1+c)$, then select $d_k = 0$, yielding $\dot x = q(t) x_k/\Delta_k$ on the entire interval as well as $\bar{q}_k = (\bar{b}/\underline{b})\underline{q}_k \le-(1+c)$.
    Integration using \eqref{eq:modified-mvt} yields $x_{k+1} = (1+q_k(\xi_k)) x_k$ with some $\xi_k \in [0,1]$.
    Noting that $q_k(\xi) \le -(1+c)$ for all $\xi \in [0,1]$, this implies
    \begin{equation}
        |x_{k+1}| = |1 + q_k(\xi_k)| |x_k| \ge c |x_k|.
    \end{equation}
    Otherwise, $\underline q_k > -(1+c)$, and thus $q_k(\xi) \ge -(1+c) {\overline b}/{\underline b}$ for all $\xi \in [0,1]$.
    In this case, $\dot x = a d_k x + q(t) x_k/\Delta_k$ provided that $|x(t)| \ge \epsilon$ holds on the interval.
    Integration yields
    \begin{equation}
        x(t) = e^{a d_k (t-t_k)} x_k + \frac{e^{a d_k (t - t_k)} - 1}{a d_k \Delta_k} q_k(\xi(t)) x_k = h(t) x_k,
    \end{equation}
    with a function $\xi : [t_k,t_{k+1}] \to [0,1]$ obtained from \eqref{eq:modified-mvt} and
    \begin{equation}
        h(t) = 1 + (e^{a d_k (t-t_k)} -1) \left(1 + \frac{q_k(\xi(t))}{a d_k \Delta_k}\right).
    \end{equation}
    Note that $h(t) \ge 1$ for all $t \in [t_k,t_{k+1}]$ provided that $d_k > \frac{(1+c) \overline b}{a \Delta_k \underline b}$, guaranteeing $|x(t)| \ge |x_k| \ge \epsilon$.
    Selecting
    \begin{equation}
        d_k = \frac{1}{a \Delta_k} \max\left\{ 2 (1+c) \frac{\overline b}{\underline b},\ln(1+2(c-1)) \right\} \le D
    \end{equation}
    and noting that $t_{k+1}-t_k = \Delta_k$ achieves
    \begin{equation}
         h(t_{k+1}) \ge 1 + (e^{a d_k \Delta_k} -1) \frac{1}{2} \ge 1 + (c-1) = c,
    \end{equation}
    yielding $|x_{k+1}| \ge c |x_k|$, concluding the proof. \qed
\end{pf}

The problem evidenced by Lemma~\ref{lem:up-bnd-f} is caused by the control action being unaware of any upper bound for the function $f$ in~\eqref{eq:sc-pert}. This justifies the fact that, under sampling, the upper bound for $f$ must be known. Assumption~\ref{as:cont-time} is thus reasonably modified as follows.
\begin{as}
    \label{as:practical}
    There exist $\underline{b},\bar{b} > 0$, and a known continuous function $\psi : \R \to \R_{\ge 0}$, such that
    \begin{align}
        \label{eq:necb}
        0 < \underline b \le b(x,t) &\le \bar b\\
        \label{eq:necf}
        |f(x,t)| &\le \psi(x),\quad \forall x\in \R, t\in\R_{\ge 0}.
    \end{align}
\end{as}
This reasonable expression for the amount of uncertainty that can be tolerated under sampling is directly related to the problem of reachability of perturbed discrete-time systems \citep{delmit_sicon69,berrho_auto71,rakker_tac06}.

\section{Sampling and Control Design for Implementation}


\subsection{Practically implementable control objective}
It is also clear that under uncertainty and a finite number of samples before the deadline, the control objective of reaching exactly a single point in the state space, e.g. the origin, is not achievable. A realistic and practically solvable control problem is then formulated as follows.
\begin{prob}
    \label{prob:practical}
  Let a minimum norm $\varepsilon>0$ for the final state, an initial time $t_0\ge 0$ and a deadline $T>0$ be given. Design a control law $u=\alpha(x,t)$ and devise a sampling strategy $t_{k+1} = S(\{x(t_i)\}_{i=0}^k,\{t_i\}_{i=0}^k) > t_k$, $k\in\{0,1,\ldots,N-1\}$, $t_N = T$, such that for every initial condition, there exists $N\in\N$ such that every trajectory satisfies
  \begin{align}
    |x(t_\ell)| \le \varepsilon
  \end{align}
  for some $\ell \in \{0,\ldots,N\}$ if the control is applied through zero-order hold as
  \begin{align}
    \label{eq:zoh-law}
    u(t) = \bar u_k = \alpha(x(t_k),t_k),\quad t\in [t_k,t_{k+1}). 
  \end{align}
\end{prob}
The design of the sampling strategy involves the selection of the sampling instants, $\{t_k\}$, where the total number of samples, $N$, may depend on the initial condition. Arguably the most practical sampling strategy is that of periodic (i.e. uniform) sampling, given by $S(\{x(t_i)\}_{i=0}^k,\{t_i\}_{i=0}^k) = t_k + \Delta$, with constant sampling period $\Delta = T/N > 0$. However, depending on the continuous-time dynamics, any such constant sampling period selection could be unable to provide global convergence, even if $b$ in~\eqref{eq:sc-pert} were known without uncertainty. This is shown next.
\begin{lem}
    \label{lem:unif-samling-notglobal}
        Consider the scalar system~\eqref{eq:sc-pert} with $b(x,t) \equiv 1$ under sampling and zero-order hold as per~\eqref{eq:zoh-law}, with a given function $\alpha : \R \times \R_{\ge 0} \to \R$ with sampling instants satisfying $t_{k+1} = t_k + \Delta$ for some fixed $\Delta > 0$ and all $k \in \{0,\ldots,N-1\}$. Let $\psi : \R \to \R_{\ge 0}$ be continuous and suppose that $\lim_{|x|\to\infty} \psi(x)/|x| = \infty$, and let $c \ge 1$. Then, there exists $M>0$ such that for all $x_0 \in \R$ there exists a function $f$ satisfying \eqref{eq:necf} such that
        $|x(t_k)| \ge M$ implies $|x(t_{k+1})| \ge c |x(t_k)|$ along the trajectory with $x(t_0) = x_0$.
%
\end{lem}
\begin{pf}
We construct a function $f$ depending on initial condition $x_0$ and function $\alpha$ as follows, by choosing on each interval $[t_k, t_{k+1})$ a constant value for $f$ depending on $x_k$ and $\bar u_k$.
Following this construction, it is seen by induction that $x_k$, $\bar u_k$ only depend on $x_0$ and $\alpha$.
    Select $M>0$ such that $\psi(x)/|x| \ge (1+2c)/\Delta$ for all $|x| \ge M$.
    If $|x_k| < M$, then set $f \equiv 0$.
    For $|x_k| \ge M$, distinguish the cases $|\bar u_k| \ge (1+c)|x_k|/\Delta$ and $|\bar u_k| < (1+c)|x_k|/\Delta$. In the first case, set $f\equiv 0$, causing $|x_{k+1}-x_k| = |\bar u_k| \Delta \ge (1+c)|x_k|$. Otherwise, set $f=\psi(x_k)\sign(x_k)$, causing $x$ to increase in absolute value since $|f|= \psi(x_k)\geq (1+2c)|x_k|/\Delta>|\bar{u}_k|$. Moreover,  $|x_{k+1}-x_k| \ge (\psi(x_k) - |\bar u_k|) \Delta \ge c|x_k|$. In any case, then $|x_{k+1}| \ge c|x_k|$.
    \qed
%
%
\end{pf}
%
%
The problem made explicit by Lemma~\ref{lem:unif-samling-notglobal} is very similar to that indicated by \citet[Theorem~3]{levant_cdc13}: the Euler discretization with fixed integration step cannot retain global fixed-time stability.

A possible solution to Problem~\ref{prob:practical} that overcomes the problem indicated by Lemma~\ref{lem:unif-samling-notglobal} is given by the following result, which allows the sampling period to be selected as a function of the initial condition.
\begin{lem}
    \label{lem:solution}
    Consider the scalar system~\eqref{eq:sc-pert} with functions $f,b$ satisfying Assumption~\ref{as:practical},
    under sampling and zero-order hold with sampling instants $t_k = t_0 + k\Delta$. Let $\phi: \mathbb{R}_{\ge 0}\to\mathbb{R}_{\geq 0}$ be a continuous strictly increasing function with $\phi(0)=0$ and $\int_0^\infty \phi(z)^{-1}dz=T$. 
    Moreover, consider 
    $$
    \alpha(x,t) = \tilde\alpha(x) = -\underline{b}^{-1}(\psi(x) + \phi(|x|))\text{sign}(x)
    $$
    in \eqref{eq:zoh-law}. 
    Then, for any $\varepsilon,M>0$, there exists a sufficiently small $\Delta>0$, such that with the sampled control actions \eqref{eq:zoh-law} and the bound $|x(t_0)| \le M$ on the initial condition,
    $|x(t_k)|\leq \varepsilon$ holds for all $k \ge T/\Delta$.
\end{lem}
\begin{rem}
\label{rem:selection-of-Delta}
By using $M = |x(t_0)|$, the sampling period $\Delta$ may be chosen as a function of the initial condition in accordance with Problem~\ref{prob:practical}.
\end{rem}
\begin{rem}
    The conditions $\phi(0)=0$ and $\int_0^\infty \phi(z)^{-1}dz=T$ imply that for all $\varepsilon>0$, $\phi$ is not Lipschitz in $[0,\varepsilon]$.
\end{rem}
\begin{pf}
All trajectories of the sampled system are solutions of the differential inclusion governed by $\dot{x}(t) \in A \psi(x(t)) + B U(x(t_k))$ for $t  \in [t_{k}, t_{k+1}]$ with $A = [-1,1]$, $B = [\underline b, \bar b]$, and
\begin{equation}
    U(x) = - \underline{b}^{-1} (\psi(x) + \phi(|x|) )\operatorname{sgn}(x)
\end{equation}
wherein $\operatorname{sgn}$ denotes the set-valued sign function defined as $\operatorname{sgn}(x) = \{ \sign(x) \}$ for $x \ne 0$ and $\operatorname{sgn}(0) = [-1,1]$.
As an auxiliary system, consider the continuous-time inclusion $\dot{\tilde x} \in \tilde G(\tilde x)$ with
\begin{equation}
    \tilde G(x) = A \psi(x) + B U(x).
\end{equation}
Set $V(\tilde x) = |\tilde x|$. For $\tilde x\neq 0$ then 
\begin{align}
\dot{V} &\in \text{sign}(\tilde x) G(\tilde x) \\
&=A \psi(\tilde x)  + [-\overline{b}/\underline{b},-1](\psi(\tilde x) + \phi(|\tilde x|)) \nonumber\\
&\subseteq [-1-\overline{b}/\underline{b},0] \psi(\tilde x) + [-\overline{b}/\underline{b},-1]\phi(|\tilde x|).
\end{align}
Since the supremum of the last set above equals $-\phi(|\tilde x|)$, then $\dot{V}\leq -\phi(V)$ if $V\neq 0$.
By the comparison lemma, there exists $s>t_0$ such that $\lim_{t\to s^-} V(\tilde x(t))=0$, 
with $s \le t_0 + \int_0^{|x(t_0)|}\phi(z)^{-1}dz < t_0 + T$ and so that $V(\tilde x(t)) = |\tilde x(t)|$ decreases for $t\in[t_0,s)$.
Consequently, $|\tilde x(t)| \le |\tilde x(t_0)| \le M$ and $\tilde x(t) = 0$ for all $t \ge t_0 + T$.
Consider now again the sampled system.
Given $\delta > 0$, define the set
\begin{equation}
G_\delta(x) = A \psi(x) + B U(x + [-\delta, \delta])
\end{equation}
and let
\begin{equation}
Q_{M,\delta} = \sup_{|x| \le M+\varepsilon}\ \sup_{\xi \in G_\delta(x)} |\xi|.
\end{equation}
If $\Delta \le \delta/Q_{M,\delta}$, then all trajectories of the sampled system satisfy the inclusion $\dot{x} \in G_{\delta}(x)$ as long as $|x(t)| \le M + \varepsilon$, because $|x(t) - x(t_{k})| \le Q_{M,\delta} \Delta \le \delta$ for $t \in [t_k, t_{k+1}]$.
It is straightforward to verify that
\begin{equation}
\label{eq:closeness-of-sets}
    G_{\delta}(x) \subset \tilde G(x + [-\delta,\delta]) + [-\delta,\delta].
\end{equation}
Hence, by \citet[Theorem 8.1]{Filippov1988}, for the given $\varepsilon$ there exists a $\delta \in (0,\varepsilon]$ such that \eqref{eq:closeness-of-sets} implies $|x(t) - \tilde x(t)| \le \varepsilon$ for all $t \in [0,T]$.
Selecting $\Delta \le \min\{ \delta/Q_{M,\delta}, T\}$ such that $N = T/\Delta$ is an integer then ensures that $|x(t_N)| \le \varepsilon$, because $\tilde x(T) = 0$.
To show that also $|x(t_k)| \le \varepsilon$ for $k > N$, note that $|x(t_{k+1})| > |x(t_k)|$ implies a sign change in $x$, i.e., $x(t_{k+1}) x(t_{k}) \le 0$, because 
\begin{equation*}
    \dot x(t) \sign(x(t)) \in ( A \psi(x(t)) + B U(x(t_k)) ) \sign(x(t)) \le 0
\end{equation*}
when $x(t) = x(t_k) \ne 0$.
Hence, either $|x(t_{k+1})|\le |x(t_k)|$ or $|x(t_{k+1})| + |x(t_k)| = |x(t_{k+1}) - x(t_k)| \le \delta \le \varepsilon$, both of which imply $|x(t_{k+1})| \le \varepsilon$.\qed
\end{pf}

\subsection{Linear control and uniform sampling}

According to Lemma~\ref{lem:unif-samling-notglobal}, a constant sampling period independent of initial conditions could be inadequate. If a bound on the possible initial conditions is known, or if the sampling period can be selected after the first state measurement becomes available, then Problem~\ref{prob:practical} can be solved with linear control and uniform sampling. For future reference, in correspondence with $\psi$ of Assumption~\ref{as:practical}, define the nonnegative nondecreasing function $\bar\psi: \R_{\ge 0} \to \R_{\ge 0}$, $r \mapsto \bar\psi_r$, as
\begin{align}
  \label{eq:barPsi-def}
  \bar\psi_r := \sup_{|x| \le r} \psi(x).
\end{align}
\begin{lem}
    Consider system~\eqref{eq:sc-pert} satisfying Assumption~\ref{as:practical} and the control objective given by Problem~\ref{prob:practical}. Let $M>0$ be given. Select $\lambda \in \R$ satisfying
    \begin{align}
        \label{eq:lam-nec-delta}
        \frac{\bar b - \underline{b}}{\bar b + \underline{b}} < \lambda < 1,
    \end{align}
    where $\underline{b},\bar{b}$ are the uncertainty bounds given by~\eqref{eq:b-disc}.
    Select the sampling period $\Delta > 0$ satisfying
    \begin{align}
        \label{eq:Delta-select}
        \Delta \le \min\left\{ \frac{\lambda - \dfrac{\bar b - \underline{b}}{\bar b + \underline{b}}}{\dfrac{\bar\psi_M}{\varepsilon}},\  \frac{T\log(1/\lambda)}{\log(M/\varepsilon)} \right\},
    \end{align}
    with $\bar\psi_M$ given by~\eqref{eq:barPsi-def}, such that the number of sampling steps before the deadline given by $N := T/\Delta$ is an integer. Select a control gain $K$ as
    \begin{align}
        \label{eq:Kselect}
        K &= \frac{\frac{1 - \lambda}{\Delta} + \bar\psi_M/\varepsilon}{\underline{b}}
    \end{align}
    Then, the linear control law and uniform sampling strategy given by
    \begin{align}
        u(t) &= -K x(t_k),\quad t\in [t_k,t_{k+1}),\\
        t_{k+1} &= t_k + \Delta = (k+1)\Delta,
    \end{align}
    steers any initial state of norm bounded by $M$ into a ball of radius $\varepsilon$ centered at the origin at some sampling instant $t_\ell \le t_N$. Moreover, if the sampling period $\Delta$ is selected small enough so that 
    \begin{align*}
        \Delta &\le \frac{\varepsilon \underline{b}}{(\underline{b}+\bar b)\bar\psi_\varepsilon}
    \end{align*}
    then the control law
    \begin{align*}
        u &= -C \sign(x(t_k)), &
        C &= \frac{\bar\psi_\varepsilon}{\underline{b}} 
    \end{align*}
    renders the closed ball $B_\varepsilon = \{x\in\R : |x| \le \varepsilon \}$ robustly invariant, i.e., if $|x(t_k)| \le \varepsilon$, then $|x(t_{k+1})| \le \varepsilon$.
\end{lem}
\begin{rem}
Analogous to Remark~\ref{rem:selection-of-Delta}, by using $M = |x(t_0)|$, the sampling period $\Delta$ and the gain $K$ may be chosen as functions of the initial condition.
\end{rem}
\begin{pf}
The closed-loop system equation is
\begin{align*}
  \dot x = f(x,t) - b(x,t) K x(t_k), \quad t\in [t_k,t_{k+1}), 
\end{align*}
with solution satisfying
\begin{align*}
  x(t) = x(t_k) + \int_{t_k}^{t} \left[ f(x(s),s) - b(x(s),s) K x(t_k)\right] ds,
\end{align*}
for $t\in [t_k,t_{k+1})$. Take absolute value to reach
\begin{align*}
  |x(t)| &= \left|x(t_k) - \int_{t_k}^{t} \left[ b(x(s),s) K x(t_k) - f(x(s),s) \right] ds\right|\\
  &= \left|x(t_k) - \int_{t_k}^{t} \left[ b(x(s),s) K - \frac{f(x(s),s)}{x(t_k)}  \right] x(t_k) ds\right|\\
  &= \left|1 - \int_{t_k}^{t} \left[ b(x(s),s) K - \frac{f(x(s),s)}{x(t_k)}  \right] ds\right| |x(t_k)| 
\end{align*}
For convergence, one could set
\begin{align}
  \label{eq:lamnec}
  \left|1 - \int_{t_k}^{t_{k+1}} \left[ b(x(s),s) K - \frac{f(x(s),s)}{x(t_k)}  \right] ds\right| \le \lambda < 1,
\end{align}
which is equivalent to
\begin{align*}
  0 < 1-\lambda \le \int_{t_k}^{t_{k+1}} \left[ b(x(s),s) K - \frac{f(x(s),s)}{x(t_k)} \right] ds \le 1+\lambda .
\end{align*}
For $|x(t_k)| \ge \varepsilon$, the expression between square brackets can be bounded as
\begin{align*}
  \underline{b}K-\frac{\bar\psi_M}{\varepsilon} \le b(x(s),s) K - \frac{f(x(s),s)}{x(t_k)} \le \bar b K + \frac{\bar\psi_M}{\varepsilon}
\end{align*}
so that if $
  1 - \lambda \le \left(\underline{b}K-\frac{\bar\psi_M}{\varepsilon}\right) \Delta$ and $\left(\bar b K + \frac{\bar\psi_M}{\varepsilon}\right) \Delta \le 1+\lambda$
then~(\ref{eq:lamnec}) would be satisfied. This is equivalent to
\begin{align}
  \label{eq:Kinterval}
  \frac{(1 - \lambda)/\Delta + \bar\psi_M/\varepsilon}{\underline{b}} \le K \le \frac{(1 + \lambda)/\Delta - \bar\psi_M/\varepsilon}{\bar{b}}
\end{align}
For a suitable value of $K$ to exist, it is necessary then that 
\begin{align}
  \label{eq:Kneccond}
  \frac{(1 + \lambda)/\Delta - \bar\psi_M/\varepsilon}{\bar{b}} - \frac{(1 - \lambda)/\Delta + \bar\psi_M/\varepsilon}{\underline{b}} \ge 0
\end{align}
or, equivalently,
\begin{align}
  (1+\lambda)\underline{b}\varepsilon - \bar\psi_M \Delta \underline{b} - (1-\lambda)\bar b \varepsilon - \bar\psi_M \Delta \bar b \notag \\
  \label{eq:Deltacond}
  = (\underline{b}-\bar b)\varepsilon + (\underline{b}+\bar b)\varepsilon \lambda - (\underline{b}+\bar b)\bar\psi_M \Delta \ge 0,
\end{align}
for which, in turn, it is necessary that
\begin{align*}
  (\underline{b}-\bar b)\varepsilon + (\underline{b}+\bar b)\varepsilon \lambda > 0
\end{align*}
which is equivalent to \eqref{eq:lam-nec-delta}.
Under~(\ref{eq:lam-nec-delta}), one can then select
\begin{align}
  \label{eq:Deltalambnd}
  0 < \Delta \le \frac{(\underline{b} + \bar b)\lambda - (\bar b - \underline{b})}{(\underline{b}+\bar b)\bar\psi_M/\varepsilon} =
  \frac{\lambda - \dfrac{\bar b - \underline{b}}{\bar b + \underline{b}}}{\dfrac{\bar\psi_M}{\varepsilon}}
\end{align}
so that~(\ref{eq:Kneccond}) is satisfied. A possible value of $K$ satisfying (\ref{eq:Kinterval}) is then given by~\eqref{eq:Kselect}.
Then, provided $|x(t_k)| \ge \varepsilon$,
\begin{align}
  \label{eq:basicconv}
  |x(t_{k+1})| &\le \lambda |x(t_k)| 
\end{align}
and
  $|x(t_k)| \le \lambda^k |x(t_0)|.$ 
For convergence from $M$ to $\varepsilon$, we need
  $|x(t_N)| \le \lambda^N M \le \varepsilon$,
giving
\begin{align}
    \label{eq:N-req}
  N \ge \frac{\log(\varepsilon/M)}{\log\lambda} = \frac{\log(M/\varepsilon)}{\log(1/\lambda)}
\end{align}
Using the fact that $T = N\Delta$, then \eqref{eq:Delta-select} is obtained from~\eqref{eq:Deltalambnd} and \eqref{eq:N-req}. All the previous design considerations will ensure~(\ref{eq:basicconv}) provided $|x(t_k)| \ge \varepsilon$, so that it will happen that $|x(t_\ell)| < \varepsilon$ for some $\ell \le N$.

For $0 < |x(t_k)| \le \varepsilon$, the specified control law
makes the solution satisfy
\begin{multline*}
  x(t_{k+1}) = \\
  x(t_k) + \int_{t_k}^{t_{k+1}} [f(x(s),s) - b(x(s),s) C\sign(x(t_k))] ds\\
  = \left(|x(t_k)| - \int_{t_k}^{t_{k+1}} \left[b(x(s),s)C - \frac{f(x(s),s)}{\sign(x(t_k))}\right] ds\right)\\
  \sign(x(t_k))
\end{multline*}
The expression between square brackets satisfies
\begin{align*}
  \underline{b} C - \bar\psi_\varepsilon \le b(x(s),s) C - \frac{f(x(s),s)}{\sign(x(t_k))} \le \bar b C + \bar\psi_\varepsilon
\end{align*}
\begin{multline*}
    \text{and then }
  -\frac{\underline{b}+\bar b}{\underline{b}} \bar\psi_\varepsilon \Delta = -(\bar b C + \bar\psi_\varepsilon) \Delta\\
  \le - \int_{t_k}^{t_{k+1}} \left[b(x(s),s)C - \frac{f(x(s),s)}{\sign(x(t_k))}\right] ds \\
  \le -(\underline{b} C - \bar\psi_\varepsilon) \Delta = 0.
\end{multline*}
\begin{multline*}
  \text{Therefore, }-\varepsilon \le |x(t_k)|- \varepsilon \le \\
  |x(t_k)| - \int_{t_k}^{t_{k+1}} \left[b(x(s),s)C - \frac{f(x(s),s)}{\sign(x(t_k))}\right] ds\\
  \le |x(t_k)| \le \varepsilon
\end{multline*}
and hence $|x(t_{k+1})| \le \varepsilon$.
If $x(t_k) = 0$, a similar reasoning shows that $|x(t_{k+1})| \le \varepsilon$ holds for any $\bar u_k \in [-C,C]$.\qed
\end{pf}

\section{Conclusions}

Although recent works have shown that prescribed-time convergence can be achieved under extreme model uncertainty, this paper shows, by considering a scalar system, that any control applied through zero-order hold could not converge or could even diverge under such uncertain conditions. Moreover, it is shown that whenever the control is applied through zero-order hold, it is necessary to not only assume bounded uncertainty but also know the uncertainty bounds in order to achieve a prescribed error bound at the deadline. Based on such necessary information, a realistic, practically implementable prescribed-convergence-time control objective is formulated and possible solutions are provided. 
Moreover, it is shown that under prior knowledge of a bound for the initial condition, or if the strategy can be selected after the first measurement becomes available, then the control objective can be achieved under uniform sampling and linear control.
Future work may involve studying the properties of sample-based implementations of higher-order algorithms based on time-varying gains that tend to infinity, as the current results suggest that the remarkable properties of continuous-time prescribed-time algorithms are significantly diminished under sampling. 


\bibliography{abbreviations}

\end{document}